\documentclass[conference]{IEEEtran}
\IEEEoverridecommandlockouts
\usepackage{cite}
\usepackage{amsmath,amssymb,amsfonts}
\usepackage{graphicx}
\usepackage{textcomp}
\usepackage{xcolor}

\usepackage{dcolumn}% Align table columns on decimal point
\usepackage{bm}% bold math
\usepackage{qcircuit}
\usepackage{braket}
\usepackage{float}
\usepackage{algorithm}
\usepackage{algorithmicx}
\usepackage{algpseudocode}
\usepackage{mathtools}
\usepackage{hyperref}
\usepackage{multirow}

\renewcommand{\figureautorefname}{Figure~\negthinspace}

\renewcommand{\sectionautorefname}{Section~\negthinspace}

\def\BibTeX{{\rm B\kern-.05em{\sc i\kern-.025em b}\kern-.08em
    T\kern-.1667em\lower.7ex\hbox{E}\kern-.125emX}}
\begin{document}

\title{An Introduction to Quantum Reinforcement Learning (QRL)
\thanks{The views expressed in this article are those of the authors and do not represent the views of Wells Fargo. This article is for informational purposes only. Nothing contained in this article should be construed as investment advice. Wells Fargo makes no express or implied warranties and expressly disclaims all legal, tax, and accounting implications related to this article.}
}

\author{\IEEEauthorblockN{Samuel Yen-Chi Chen}
\IEEEauthorblockA{
% \textit{dept. name of organization (of Aff.)} \\
\textit{Wells Fargo}\\
New York, NY, USA \\
yen-chi.chen@wellsfargo.com}
}

\maketitle

\begin{abstract}
Recent advancements in quantum computing (QC) and machine learning (ML) have sparked considerable interest in the integration of these two cutting-edge fields. Among the various ML techniques, reinforcement learning (RL) stands out for its ability to address complex sequential decision-making problems. RL has already demonstrated substantial success in the classical ML community. Now, the emerging field of Quantum Reinforcement Learning (QRL) seeks to enhance RL algorithms by incorporating principles from quantum computing. This paper offers an introduction to this exciting area for the broader AI and ML community.
\end{abstract}


\section{Introduction}
Quantum computing (QC) offers the potential for substantial computational advantages in specific problems compared to classical computers \cite{nielsen2010quantum}. Despite the current limitations of quantum devices, such as noise and imperfections, significant efforts are being made to achieve quantum advantages. One prominent area of focus is quantum machine learning (QML), which leverages quantum computing principles to enhance machine learning tasks. Most QML algorithms rely on a hybrid quantum-classical paradigm, which divides the computational task into two components: quantum computers handle the parts that benefit from quantum computation, while classical computers process the parts they excel at.

Variational quantum algorithms (VQAs) \cite{bharti2022noisy} form the foundation of current quantum machine learning (QML) approaches. QML has demonstrated success in various machine learning tasks, including classification \cite{mitarai2018quantum,chen2021end,qi2023qtnvqc,chen2022quantumCNN}, sequential learning \cite{chen2022quantumLSTM,chen2022reservoir}, natural language processing \cite{li2023pqlm,yang2022bert,di2022dawn,stein2023applying}, and reinforcement learning \cite{chen19,chen2022variationalQRL,lockwood2020reinforcement,skolik2021quantum,chen2023quantumLSTM_RL,meyer2022survey,chen2024efficient}. Among these areas, quantum reinforcement learning (QRL) is an emerging field where researchers are exploring the application of quantum computing principles to enhance the performance of reinforcement learning agents. This article provides an introduction to the concepts and recent developments in QRL.
\section{Quantum Neural Networks}
\subsection{Quantum Computing}
A \emph{qubit} represents the fundamental unit of quantum information processing. Unlike a classical \emph{bit}, which is restricted to holding a state of either $0$ or $1$, a qubit can simultaneously encapsulate the information of both $0$ and $1$ due to the principle of superposition.
A single qubit quantum state can be expressed as $\ket{\Psi} = \alpha\ket{0} + \beta\ket{1}$, where $\ket{0} = [1, 0]^{T}$ and $\ket{1} = [0, 1]^{T}$ are column vectors, and $\alpha$ and $\beta$ are complex numbers. In an $n$-qubit system, the state vector has a length of $2^{n}$.
\emph{Quantum gates} $U$ are utilized to transform a quantum state, represented as $\ket{\Psi}$, to another state $\ket{\Psi'}$ through the operation $\ket{\Psi'} = U \ket{\Psi}$. These quantum gates are \emph{unitary transformations} that satisfy the condition $U U^{\dagger} = U^{\dagger} U = \mathbb{I}_{2^{n} \times 2^{n}}$, where $n$ denotes the number of qubits. It has been demonstrated that a small set of basic quantum gates is sufficient for universal quantum computation. One such set includes single-qubit gates $H$, $\sigma_x$, $\sigma_y$, $\sigma_z$, $R_{x}(\theta) = e^{-i\theta\sigma_x/2}$, $R_{y}(\theta) = e^{-i\theta\sigma_y/2}$, $R_{z}(\theta) = e^{-i\theta\sigma_z/2}$, and the two-qubit gate CNOT. In quantum machine learning (QML), rotation gates $R_{x}$, $R_{y}$, and $R_{z}$ are particularly crucial as their rotation angles can be treated as trainable or learnable parameters subject to optimization. For quantum operations on multi-qubit systems, the unitary transformation can be constructed via the tensor product of individual single-qubit or two-qubit operations, $U = U_{1} \otimes U_{2} \otimes \cdots \otimes U_{k}$.
At the final stage of a quantum circuit, a procedure known as \emph{measurement} is performed. A single execution of a quantum circuit generates a binary string. This procedure can be repeated multiple times to determine the probabilities of different computational bases (e.g., $\ket{0,\cdots,0}$, $\cdots$, $\ket{1,\cdots,1}$) or to calculate expectation values (e.g., Pauli $X$, $Y$, and $Z$).
\subsection{Variational Quantum Circuits}
\begin{figure}[htbp]
%\vskip -0.2in
\vskip -0.15in
\begin{center}
\includegraphics[width=1\columnwidth]{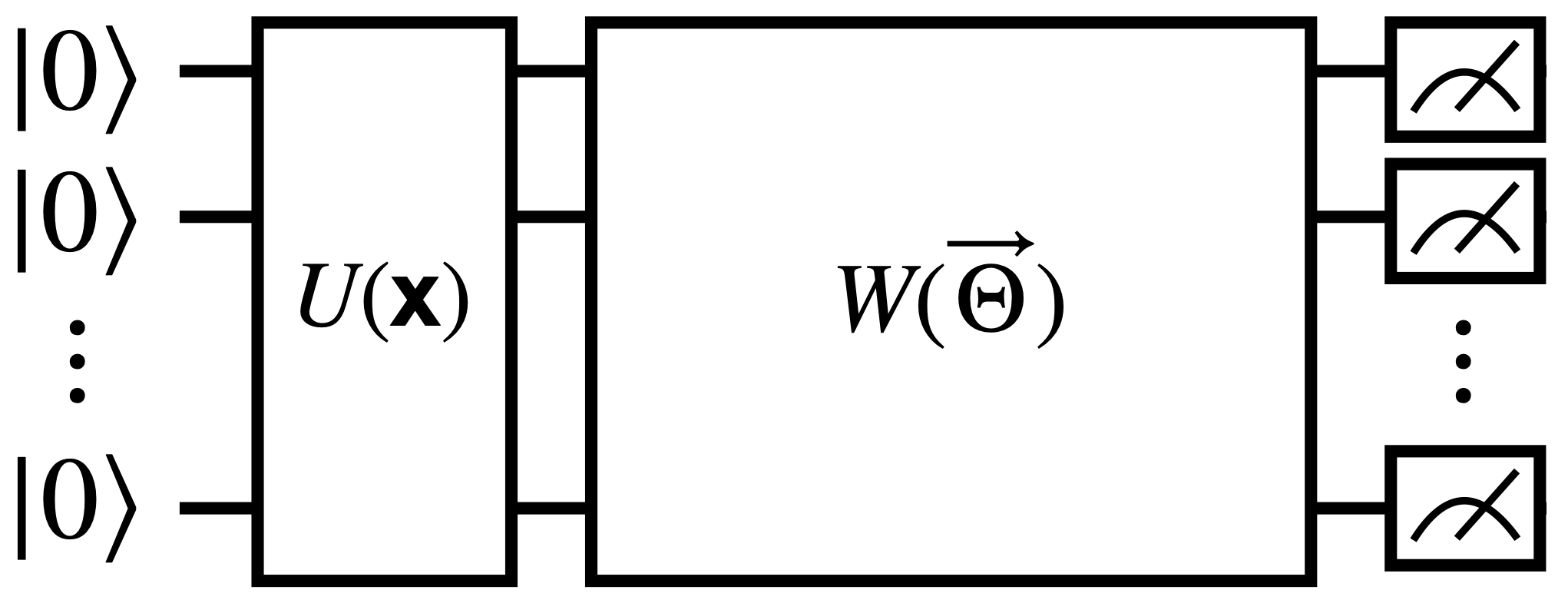}\vskip -0.1in
\caption{{\bfseries Generic Structure of a Variational Quantum Circuit (VQC).}}
\label{fig:generic_VQC}
\end{center}
\vskip -0.15in
\end{figure}

\emph{Variational quantum circuits} (VQCs), also referred to as \emph{parameterized quantum circuits} (PQCs), represent a specialized class of quantum circuits with trainable parameters. VQCs are extensively utilized within the current hybrid quantum-classical computing framework \cite{bharti2022noisy} and have demonstrated specific types of quantum advantages \cite{abbas2021power,caro2022generalization,du2020expressive}.
There are three fundamental components in a VQC: \emph{encoding circuit}, \emph{variational circuit} and the final \emph{measurements}.
As shown in \figureautorefname{\ref{fig:generic_VQC}}, the encoding circuit $U(\mathbf{x})$ transforms the initial quantum state $\ket{0}^{\otimes n}$ into $\ket{\Psi} = U(\mathbf{x})\ket{0}^{\otimes n}$. Here the $n$ represents the number of qubits, $\ket{0}^{\otimes n}$ represents the $n$-qubit initial state $\ket{0, \cdots, 0}$ and the $U(\mathbf{x})$ represents the unitary which depends on the input value $\mathbf{x}$. 
The measurement process extracts data from the VQC by assessing either a subset or all of the qubits, producing a classical bit sequence for further use.
Running the circuit once yields a bit sequence such as "0,0,1,1." However, preparing and executing the circuit multiple times (shots) generates expectation values for each qubit.
Most works mentioned in this survey focus on the evaluation of Pauli-$Z$ expectation values derived from measurements in VQCs.
Generally, the mathematical expression of the VQC can be expressed as $\overrightarrow{f(\mathbf{x} ; \Theta)}=\left(\left\langle\hat{Z}_1\right\rangle, \cdots,\left\langle\hat{Z}_n\right\rangle\right)$ , where $\left\langle\hat{Z}_{k}\right\rangle =\left\langle 0\left|U^{\dagger}(\mathbf{x})W^{\dagger}(\Theta) \hat{Z_{k}} W(\Theta)U(\mathbf{x})\right| 0\right\rangle$.
In the hybrid quantum-classical framework, the VQC can be integrated with other classical components, such as deep neural networks and tensor networks, or with other quantum components, including additional VQCs. The entire model can be optimized in an end-to-end manner using either gradient-based \cite{chen2021end,qi2023qtnvqc} or gradient-free \cite{chen2022variationalQRL} methods. For gradient-based methods like gradient descent, the gradients of quantum components can be computed via the parameter-shift rules \cite{mitarai2018quantum,schuld2019evaluating,bergholm2018pennylane}.
\section{Quantum Reinforcement Learning}
\subsection{Reinforcement Learning}
\emph{Reinforcement Learning} (RL) is a pivotal paradigm within machine learning, where an autonomous entity known as the \emph{agent} learns to make decisions through iterative interactions with its environment~\cite{sutton2018reinforcement}. The agent operates within a defined \emph{environment}, represented as $\mathcal{E}$, over discrete time steps. At each time step $t$, the agent receives \emph{state} or \emph{observation} information, denoted as $s_t$, from the environment $\mathcal{E}$. Based on this information, the agent selects an \emph{action} $a_t$ from a set of permissible actions $\mathcal{A}$, guided by its \emph{policy} $\pi$. The policy $\pi$ acts as a function that maps the current state or observation $s_t$ to the corresponding action $a_t$. Notably, the policy can be stochastic, indicating that for a given state $s_t$, the action $a_t$ is determined by a probability distribution $\pi(a_t|s_t)$.

\begin{figure}[htbp]
\centering
\vskip -0.15in
\includegraphics[width=1\linewidth]{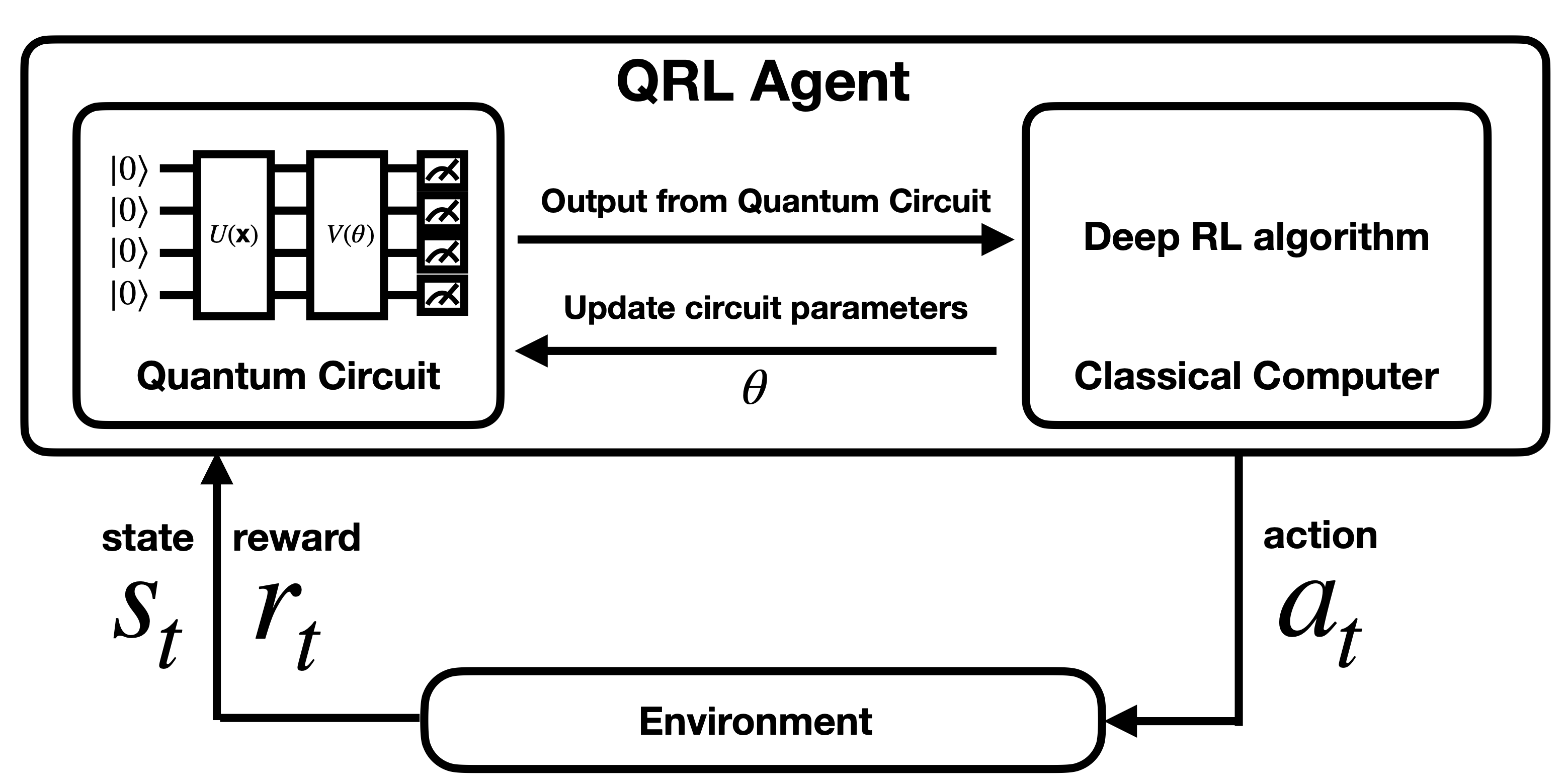}% Here is how to import EPS art
\caption{{\bfseries Concept of quantum reinforcement learning (QRL). }}
\label{fig:QRL_Concept}
\vskip -0.1in
\end{figure}

Upon executing action $a_t$, the agent transitions to the subsequent state $s_{t+1}$ and receives a scalar \emph{reward} $r_t$. This cycle continues until the agent reaches a terminal state or fulfills a specified stopping condition, such as a maximum number of steps. We define an \emph{episode} as the sequence beginning from an initial state, following the described process, and concluding either at the terminal state or upon meeting the stopping criterion. The use of quantum neural networks for learning policy or value functions is referred to as quantum reinforcement learning (QRL).
The idea of QRL is illustrated in \figureautorefname{\ref{fig:QRL_Concept}}.
For a comprehensive review of current QRL domain, refer to the review article \cite{meyer2022survey}.
\subsection{Quantum Deep Q-learning}
\begin{figure}[htbp]
\centering
\vskip -0.16in
\includegraphics[width=1\linewidth]{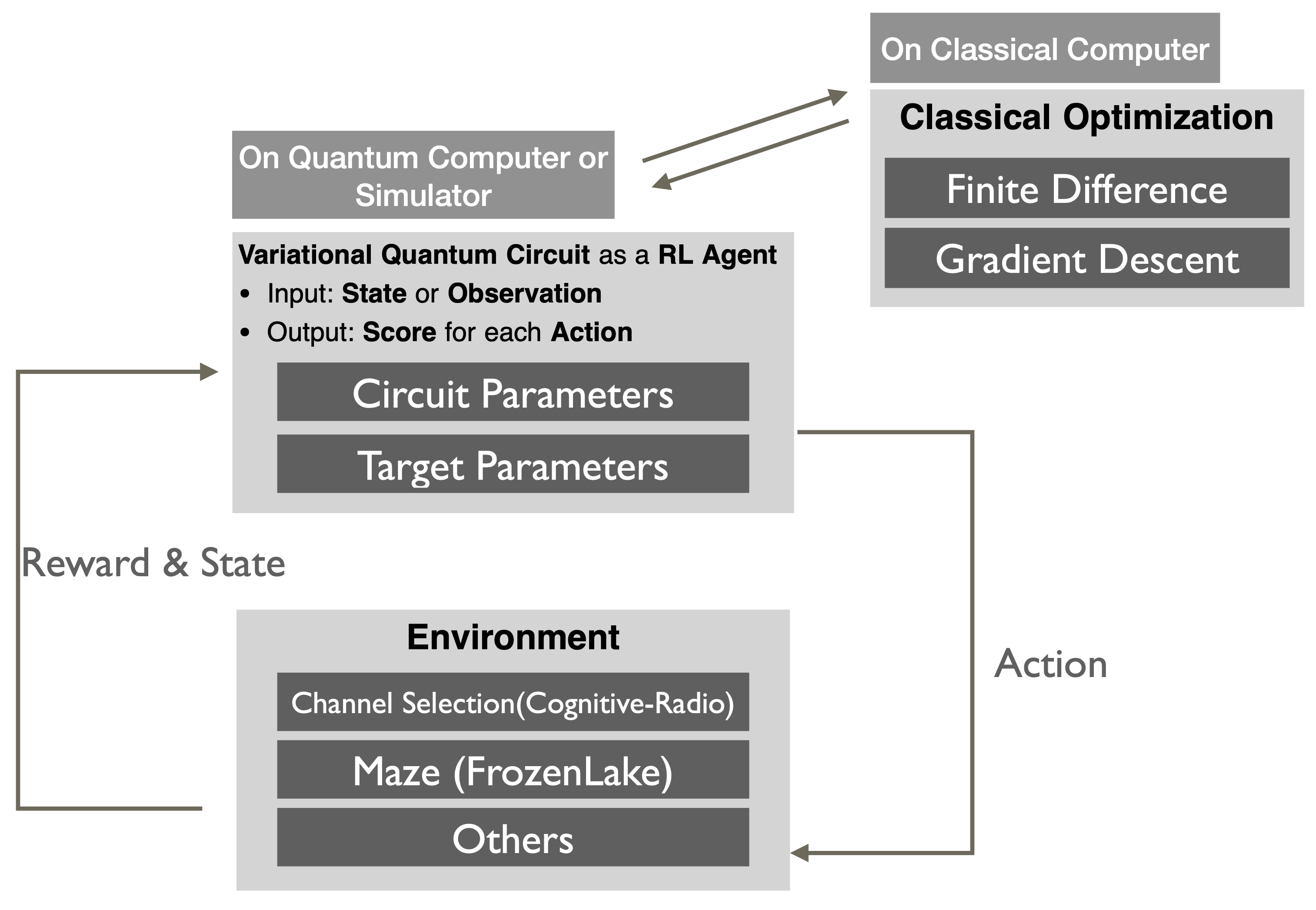}% Here is how to import EPS art
\caption{{\bfseries Quantum deep Q-learning. }}
\label{fig:VQDQN_scheme}
\vskip -0.16in
\end{figure}

$Q$-learning \cite{sutton2018reinforcement} is a fundamental model-free RL algorithm. It learns the optimal action-value function and operates \emph{off-policy}. The process begins with the random initialization of $Q^{\pi}(s,a)$ for all states $s \in S$ and actions $a \in \mathcal{A}$, stored in a $Q$-table. The $Q^{\pi}(s,a)$ estimates are updated using the Bellman equation:
\begin{align}
  Q\left(s_{t}, a_{t}\right) & \leftarrow  Q\left(s_{t}, a_{t}\right) \nonumber \\
  &+\alpha\left[r_{t}+\gamma \max _{a} Q\left(s_{t+1}, a\right)-Q\left(s_{t}, a_{t}\right)\right].
\end{align}
The conventional $Q$-learning approach offers the optimal action-value function but is impractical for problems requiring extensive memory, especially with high-dimensional state ($s$) or action ($a$) spaces. In environments with continuous states, storing $Q(s,a)$ in a table is inefficient or impossible. To address this challenge, neural networks (NNs) are used to represent $Q^{\pi}(s,a),  \forall s \in S, a \in \mathcal{A}$, leading to \emph{deep $Q$-learning}. The network in this technique is known as a deep $Q$-network (DQN) \cite{mnih2015human}.
To enhance the stability of DQN, techniques such as \emph{experience replay} and the use of a \emph{target network} are employed \cite{mnih2015human}. Experience replay stores \emph{experiences} as transition tuples ${s_{t}, a_{t}, r_{t}, s_{t+1}}$ in a memory or buffer. After gathering sufficient experiences, the agent randomly samples a batch to compute the loss and update DQN parameters. Additionally, to reduce correlation between target and prediction, a \emph{target network}, which is a duplicate of the DQN, is used. The DQN parameters $\theta$ are updated iteratively, while the target network parameters $\theta^{-}$ are updated periodically.
The DQN training is done via minimizing the mean square error (MSE) loss function:
\begin{equation}
\resizebox{0.9\columnwidth}{!}{
$L(\theta)=\mathbb{E}\left[\left(r_{t}+\gamma \max _{a^{\prime}} Q\left(s_{t+1}, a^{\prime} ; \theta^{-}\right)-Q\left(s_{t}, a_{t} ; \theta\right)\right)^{2}\right]$}
\end{equation}
Other loss functions such as Huber loss or mean absolute error (MAE) can also be used.
The first VQC-based QRL is described in the work \cite{chen19} in which a VQC is designed to solve environments with discrete observations such as the Frozen Lake and Cognitive-Radio. The design follows the original idea in classical deep $Q$-learning \cite{mnih2015human}. As shown in \figureautorefname{\ref{fig:VQDQN_scheme}}, there are target network and experience replay in this quantum DQN. They are basically two sets of quantum circuit parameters. The quantum agent is optimized via gradient descent algorithm such as RMSProp in the hybrid quantum-classical manner.
Later, more sophisticated efforts in the area of quantum DQN take into account continuous observation spaces like Cart-Pole \cite{lockwood2020reinforcement,skolik2021quantum}.
\subsection{Quantum Policy Gradient Methods}
In contrast to \emph{value-based} RL algorithms, such as $Q$-learning, which depend on learning a value function to guide decision-making at each time step, \emph{policy gradient} methods aim to optimize a policy function directly. This policy function $\pi(a|s;\theta)$ is parameterized by $\theta$. The parameters $\theta$ are adjusted using gradient ascent on the expected total return, $\mathbb{E}[R_{t}]$. A prominent example of a policy gradient algorithm is the REINFORCE algorithm \cite{williams1992simple}.
The policy function $\pi(a|s;\theta)$ can be implemented using a VQC, where the rotation parameters serve as $\theta$. In \cite{jerbi2021variational}, the authors employ the REINFORCE algorithm to train a VQC-based policy. Their results demonstrate that VQC-based policies can achieve performance comparable to or exceeding that of classical DNNs on several standard benchmarks.
In the traditional REINFORCE algorithm, parameter updates for $\theta$ are based on the gradient $\nabla_{\theta} \log \pi\left(a_{t} | s_{t} ; \theta\right) R_{t}$, which provides an unbiased estimate of $\nabla_{\theta} \mathbb{E}\left[R_{t}\right]$. However, this gradient estimate can exhibit high variance, which may lead to difficulties or instability during training. To address this issue and reduce variance while preserving unbiasedness, a \emph{baseline} term can be subtracted from the return. This baseline, $b_{t}(s_{t})$, is a learned function of the state $s_{t}$. The update rule then becomes $\nabla_{\theta} \log \pi\left(a_{t} | s_{t} ; \theta\right)\left(R_{t}-b_{t}\left(s_{t}\right)\right)$.
A typical choice for the baseline $b_t(s_t)$ in RL is an estimate of the value function $V^\pi(s_t)$. Employing this baseline generally leads to a reduction in the variance of the policy gradient estimate \cite{sutton2018reinforcement}. The term $R_t - b_t = Q(s_t, a_t) - V(s_t)$ represents the \emph{advantage} $A(s_t, a_t)$ of taking action $a_t$ in state $s_t$. This advantage can be viewed as a measure of how favorable or unfavorable action $a_t$ is compared to the average value of the state $s_t$. This method is referred to as the advantage actor-critic (A2C) approach, where the policy $\pi$ serves as the \emph{actor} and the value function $V$ acts as the \emph{critic} \cite{sutton2018reinforcement}.
Similar to traditional policy gradient methods, the A2C algorithm can be implemented using VQCs. In \cite{kolle2024quantum}, the authors utilize VQCs to construct both the actor (policy function) and the critic (value function). Their study demonstrates that, for comparable numbers of model parameters, a hybrid approach—where classical neural networks post-process the outputs from the VQC—achieves superior performance across the tested environments.
The asynchronous advantage actor-critic (A3C) algorithm \cite{mnih2016asynchronous} is an enhanced variant of the A2C method that utilizes multiple concurrent actors to learn the policy through parallel processing. This approach involves deploying several agents across several instances of the environment, enabling them to experience a wide range of states simultaneously. By reducing the correlation between states or observations, this method improves the numerical stability of on-policy RL algorithms like actor-critic \cite{mnih2016asynchronous}. Moreover, asynchronous training eliminates the need for extensive replay memory, which helps in reducing memory usage \cite{mnih2016asynchronous}. A3C achieves high sample efficiency and robust learning performance, making it a favored choice in RL. In the context of quantum RL, asynchronous or distributed training can further boost sampling efficiency and leverage the capabilities of multiple quantum computers or quantum processing units (QPUs). In \cite{CHEN2023321Async}, the authors extend the A3C framework to quantum settings, showing that VQC actors and critics can outperform classical models when the sizes of the models are comparable.
\subsection{Quantum RL with Evolutionary Optimization}

\begin{figure}[htbp]
\centering
\vskip -0.14in
\includegraphics[width=1\linewidth]{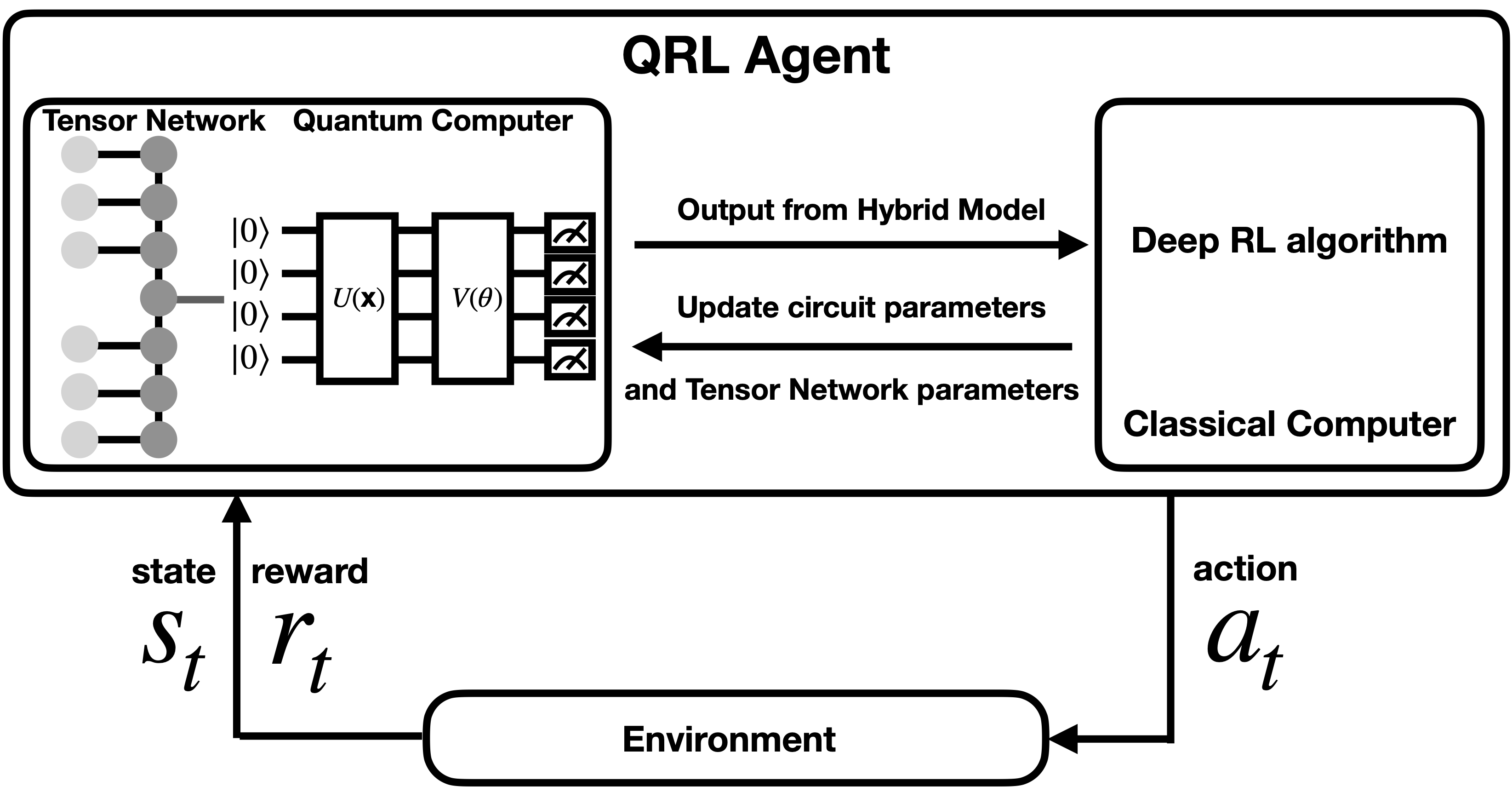}% Here is how to import EPS art
\caption{{\bfseries Hybrid Quantum-Classical RL with Tensor Networks. }}
\label{fig:TN_QRL_Concept}
\vskip -0.16in
\end{figure}

One of the significant challenges in current QML applications is the limitation of quantum computers or quantum simulation software in processing input dimensions. These systems can only handle inputs up to a certain level, which is insufficient for encoding larger vectors. In QRL, this constraint means the observation vector that the quantum agent can process from the environment is severely restricted. To address this issue, various dimensional reduction methods have been proposed. Among these, a hybrid quantum-classical approach that incorporates a classical learnable model with a VQC has shown promising results. In the work by Chen et al. \cite{chen2022variationalQRL}, a quantum-inspired classical model based on a specific type of tensor network, known as a matrix product state (MPS), is integrated with a VQC to function as a learnable compressor \cite{chen2021end} (see \figureautorefname{\ref{fig:TN_QRL_Concept}}). The hybrid architecture MPS-VQC, including the tensor network and VQC, is randomly initialized, and the entire model is trained in an end-to-end manner.
Although gradient-based methods have achieved considerable success in RL, several challenges remain. Notably, these methods can become trapped in local minima or fail to converge to the optimal solution, particularly in sparse RL environments where the agent frequently receives zero rewards during episodes.
Evolutionary optimization techniques have been proposed to address these challenges in classical RL and have demonstrated significant success \cite{such2017deep}. A similar approach can be applied to hybrid quantum-classical RL models. Specifically, a population $\mathcal{P}$ of $N$ agents, represented as parameter vectors $\Theta_{i}, i \in {1, \cdots, N}$, is randomly initialized. In each \emph{generation}, the top-performing agents are selected to serve as \emph{parents} for generating the next generation of agents/parameter vectors. The update rules for the new parameters involve adding Gaussian noise to the parent parameters. This method has been shown to optimize MPS-VQC models effectively and to outperform NN-VQC in selected benchmarks \cite{chen2022variationalQRL}.
\subsection{Quantum RL with Recurrent Policies}
The previously mentioned quantum RL methods primarily utilize various VQCs without incorporating recurrent structures. However, recurrent connections are essential in classical machine learning for retaining memory of past time steps. Certain RL tasks necessitate that agents have the capability to remember information from previous time steps to select optimal actions. For instance, environments with partial observability often require agents to make decisions based not only on information from the current time step but also on information accumulated from the past.
In classical ML, recurrent neural networks (RNNs), such as long short-term memory (LSTM) \cite{hochreiter1997long}, have been proposed to solve tasks with temporal dependencies. The quantum version of LSTM (QLSTM) has been designed by replacing classical neural networks with VQCs \cite{chen2022quantumLSTM}. It has been shown that QLSTM can outperform classical LSTM in several time-series prediction tasks when the model sizes are similar \cite{chen2022quantumLSTM}.
To address RL environments with partial observability or those requiring temporal memories, QRL agents utilizing QLSTM as the value or policy function have been proposed in \cite{chen2023quantumLSTM_RL}. It has been demonstrated that QLSTM-based value or policy functions enable QRL agents to outperform classical LSTM models with a similar number of parameters.

While the QLSTM-based models achieve significant results in several benchmarks, there is at least one major challenge preventing such models from wide applications. The training of RNNs, both in quantum and classical, requires significant computational resources due to the requirement of performing \emph{backpropagation-through-time} (BPTT).
One might question whether it is possible to leverage the capabilities of QLSTM without the need for gradient calculations with respect to the quantum parameters. Indeed, it has been demonstrated that a randomly initialized RNN can function as a \emph{reservoir}, transforming input information into a high-dimensional space. The only part that requires training is the linear layer following the reservoir.
Quantum RNNs, such as QLSTM, can also be utilized as reservoirs \cite{chen2022reservoir}. It has been shown that even without training, the QLSTM reservoir can achieve performance comparable to fully trained models \cite{chen2022reservoir}.
To further enhance the performance of QLSTM-based QRL agents and reduce training resource requirements, a randomly initialized QLSTM can be employed as a reservoir in an RL agent \cite{chen2024efficient}. Numerical simulations have demonstrated that the QLSTM reservoir can achieve performance comparable to, and sometimes superior to, fully trained QLSTM RL agents.
\subsection{Quantum RL with Fast Weight Programmers}
An alternative approach for developing a QRL model that can memorize temporal or sequential dependencies without utilizing quantum RNNs is the \emph{Quantum Fast Weight Programmers} (QFWP). The idea of \emph{Fast Weight Programmers} (FWP) was originally proposed in the work of Schmidhuber \cite{schmidhuber1992learning,schmidhuber1993reducing}.
In this sequential learning model, two distinct neural networks (NN) are utilized: the \emph{slow programmer} and the \emph{fast programmer}. Here, the NN weights act as the model/agent's \emph{program}. The core concept of FWP involves the slow programmer generating \emph{updates} or \emph{changes} to the fast programmer's NN weights based on observations at each time-step.
This \emph{reprogramming} process quickly redirects the fast programmer's attention to salient information within the incoming data stream. Notably, the slow programmer does not completely overwrite the fast programmer but instead applies updates or changes. This approach allows the fast programmer to incorporate previous observations, enabling a simple feed-forward NN to manage sequential prediction or control without the high computational demands of recurrent neural networks (RNNs).
The idea of FWP can be further extended into the hybrid quantum-classical regime as described in the work \cite{chen2024learning}.
In the work \cite{chen2024learning}, classical neural networks are used to construct the \emph{slow} networks, which generate values to update the parameters of the \emph{fast} networks, implemented as a VQC.
\begin{figure}[htbp]
% \vskip -0.1in
\centering
\includegraphics[width=1\linewidth]{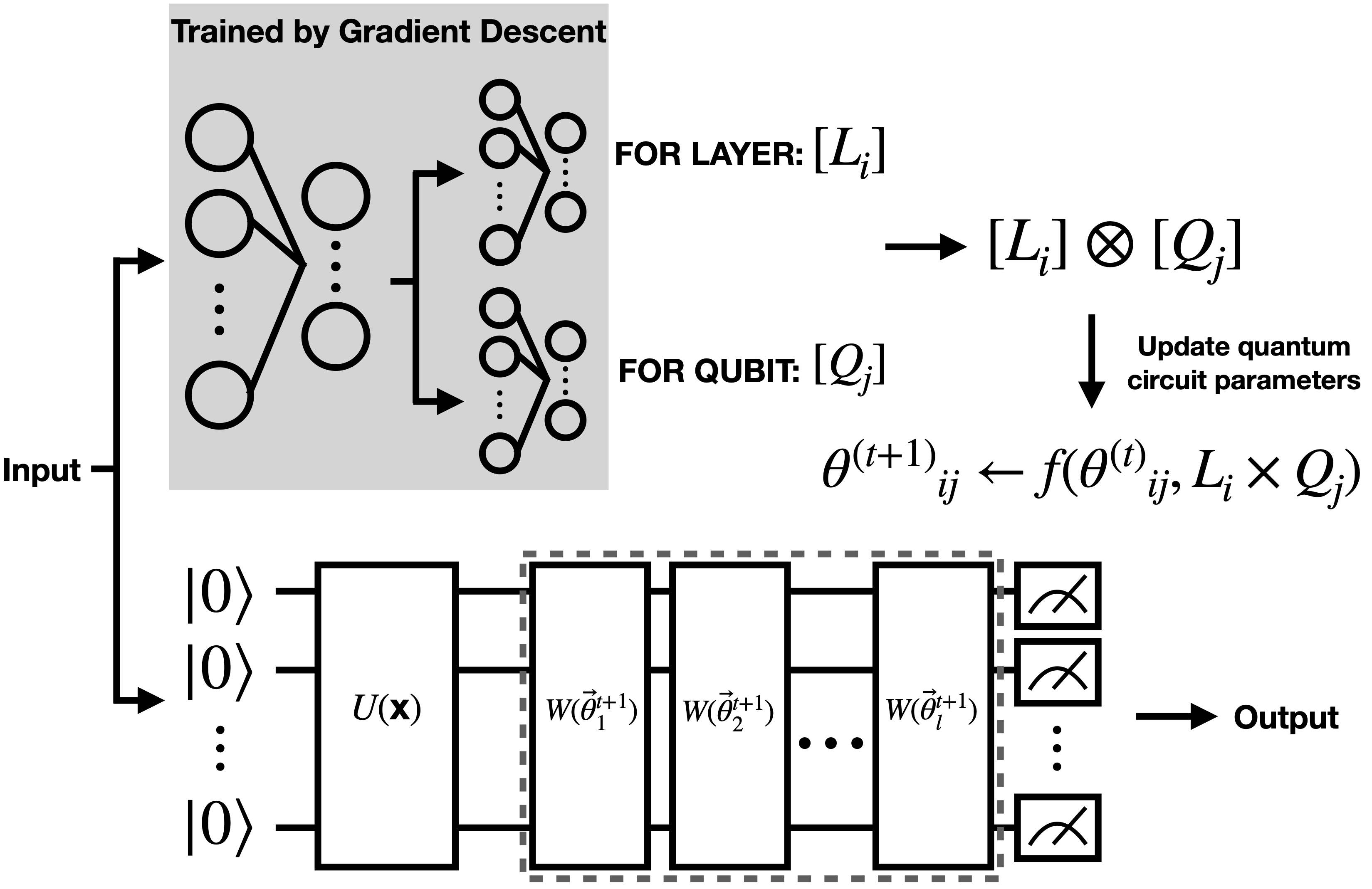}% Here is how to import EPS art
\caption{{\bfseries Quantum Fast Weight Programmers. }}
\label{fig:vqFWP_Concept}
\vskip -0.16in
\end{figure}

As illustrated in \figureautorefname{\ref{fig:vqFWP_Concept}}, the input vector $\Vec{x}$ is first processed by a classical neural network encoder. The encoder's output is then fed into two additional neural networks. One network generates an output vector $[L_{i}]$ corresponding to the number of VQC layers, while the other produces an output vector $[Q_{j}]$ matching the number of qubits in the VQC.
We then calculate the outer product of $[L_{i}]$ and $[Q_{j}]$. It can be written as $[L_{i}] \otimes [Q_{j}] = [M_{ij}] = [L_{i} \times Q_{j}] = 
\begin{bmatrix}
L_{1} \times Q_{1} & L_{1} \times Q_{2} & \cdots & L_{1} \times Q_{n}\\
L_{2} \times Q_{1} & L_{2} \times Q_{2} & \cdots & L_{2} \times Q_{n}\\
\vdots             &       \ddots       &        &        \vdots     \\
L_{l} \times Q_{1} & L_{l} \times Q_{2} & \cdots & L_{l} \times Q_{n}\\
\end{bmatrix}$, where $l$ is the number of learnable layers in VQC and $n$ is the number of qubits.
At time $t+1$, the updated VQC parameters can be calculated as $\theta^{t+1}_{ij} = f(\theta^{t}_{ij}, L_{i} \times Q_{j})$, where $f$ combines the previous parameters $\theta^{t}_{ij}$ with the newly computed $L_{i} \times Q_{j}$.
In the time series modeling and RL tasks in \cite{chen2024learning}, the \emph{additive} update rule is used. The new circuit parameters are calculated as $\theta^{t+1}_{ij} = \theta^{t}_{ij} + L_{i} \times Q_{j}$. This method preserves information from previous time steps in the circuit parameters, influencing the VQC behavior with each new input $\Vec{x}$.
The output from the VQC can be further processed by components such as scaling, translation, or a classical neural network to refine the final results.
\subsection{\label{sec:QRL_with_QAS}Quantum RL with Quantum Architecture Search}
\begin{figure}[htbp]
\centering
\vskip -0.1in
\includegraphics[width=1\linewidth]{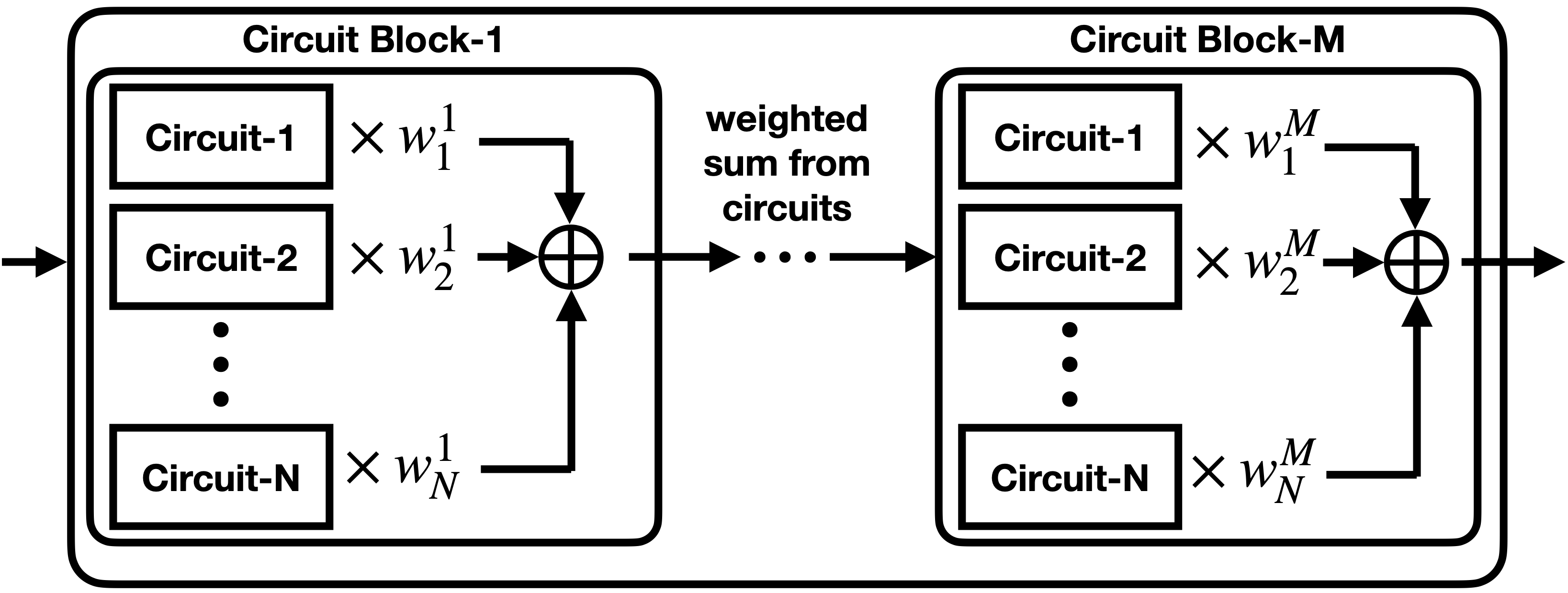}% Here is how to import EPS art
\caption{{\bfseries Differentiable Quantum Architecture Search (DiffQAS). }}
\label{fig:DiffQAS_Scheme}
\vskip -0.16in
\end{figure}

While QRL has demonstrated effectiveness across various problem domains, the design of successful architectures is far from trivial. Developing VQC architectures tailored to specific problems requires substantial effort and expertise.
The field of \emph{quantum architecture search} (QAS) focuses on developing methods to identify high-performing quantum circuits for specific tasks. A QAS problem is formulated by specifying a particular goal (e.g., total returns in RL) and the constraints of the quantum device (e.g., maximum number of quantum operations, set of allowed quantum gates).
QAS has been explored in the context of QRL. For instance, in \cite{ding2022evolutionary}, evolutionary algorithms are employed to search for high-performing circuits. The authors define a set of candidate VQC blocks, including entangling blocks, data-encoding blocks, variational blocks, and measurement blocks. The objective of the evolutionary search is to determine an optimal sequence of these blocks, given a constraint on the maximum number of circuit blocks. While this approach has shown effectiveness in the evaluated cases, scalability issues may arise as the search space expands.
Differentiable quantum architecture search (DiffQAS) methods, as proposed in \cite{zhang2022differentiable}, draw inspiration from differentiable neural architecture search in classical deep learning to identify effective quantum circuits for RL. In \cite{sun2023differentiable}, the authors apply DiffQAS to quantum deep $Q$-learning. They parameterize a probability distribution $P(k, \alpha)$ for circuit architecture $k$ using $\alpha$. During training, mini-batches of VQCs are sampled, and the weighted loss is calculated based on the distribution $P(k, \alpha)$. Both the architecture parameter $\alpha$ and the quantum circuit parameters $\theta$ are updated using conventional gradient-based methods.
In \cite{chen2024differentiable}, the authors extend the DiffQAS framework to asynchronous QRL. This extension allows multiple parallel instances (a single instance is shown in \figureautorefname{\ref{fig:DiffQAS_Scheme}
}) to optimize their own structural weights (denoted as $w$ in \figureautorefname{\ref{fig:DiffQAS_Scheme}
}) alongside the VQC parameters. The gradients of these structural weights and quantum circuit parameters are shared across instances to enhance the training process.
\section{Quantum RL Applications and Challenges}

QRL can be extended to multi-agent settings and applied in fields like wireless communication and autonomous control systems \cite{park2023quantumMACN}. Additionally, as discussed in \sectionautorefname{\ref{sec:QRL_with_QAS}}, QAS involves sequential decision-making and can be addressed through RL. In \cite{chen2023quantumRL_QAS}, a QRL approach is developed to discover quantum circuit architectures that generate desired quantum states.
In the NISQ era, a major challenge for QML applications is the limited quantum resources, which complicates both the training and inference phases.
In \cite{liu2024quantumTrain,liu2024qtrl}, the authors propose a method using a QNN to generate classical NN weights. For an $N$-qubit QNN, measuring the expectation values of individual qubits provides up to $N$ values. However, collecting the probabilities of all computational basis states ${\ket{00 \cdots 0}, \cdots, \ket{11 \cdots 1}}$ yields $2^{N}$ values. These values can be rescaled and used as NN weights. Thus, for an NN with $M$ weights, only $\lceil \log_{2}M \rceil$ qubits are needed to generate the weights. Numerical simulations demonstrate that the quantum circuit can efficiently generate NN weights, achieving inference performance comparable to conventional training.
Future research could further explore the trainability challenges in QRL models highlighted by Sequeira et al. \cite{sequeira2024trainability}, which are key to enhancing their practical performance.
\section{Conclusion and Outlook}
This paper introduces the concept of quantum reinforcement learning (QRL), where variational quantum circuits (VQCs) are used as policy and value functions. It also explores advanced constructs, including quantum recurrent policies, quantum fast weight programmers, and QRL with differentiable quantum architectures. QRL holds the potential to offer quantum advantages in various sequential decision-making tasks.

\bibliographystyle{IEEEtran}
\bibliography{bib/classical_rl,bib/vqc,bib/qml_examples,bib/tool,bib/qrl,bib/classical_ml,bib/fwp,bib/qas,bib/qc,bib/qt}
\end{document}